\documentclass[epj,final]{svjour}
\usepackage{latexsym,amsmath,amssymb}
\usepackage{graphics,graphicx}
\usepackage[latin1]{inputenc}

\newcommand{\mgb}{MgB$_2$}

\begin{document}

\title{Microwave surface resistance of pristine and neutron-irradiated MgB$_2$ samples in magnetic field}

\author{M. Bonura \inst{1} \and A. Agliolo~Gallitto \inst{1} \and M. Li~Vigni \inst{1}
\and C. Ferdeghini \inst{2} \and C. Tarantini \inst{2}}

\institute{CNISM and Dipartimento di Scienze Fisiche e Astronomiche,
Università di Palermo, Via Archirafi 36, I-90123 Palermo, Italy%
\and
CNR-INFM-LAMIA and Dipartimento di Fisica, Università di Genova, Via
Dodecaneso 33, I-16146 Genova, Italy}

%\date{\today}
% The correct dates will be entered by Springer

\abstract{We report on the microwave surface resistance of two
polycrystalline $\mathrm{Mg}^{11} \mathrm{B}_2$ samples; one consists of
pristine material, the other has been irradiated at very high neutron
fluence. It has already been reported that in the strongly irradiated
sample the two gaps merge into a single value. The mw surface resistance
has been measured in the linear regime as a function of the temperature and the DC magnetic
field, at increasing and decreasing fields. The results obtained in the
strongly irradiated sample are quite well justified in the framework of a
generalized Coffey and Clem model, in which we take into account the field
distribution inside the sample due to the critical state. The results
obtained in the pristine sample show several anomalies, especially at low
temperatures, which cannot be justified in the framework of standard
models for the fluxon dynamics. Only at temperatures near $T_c$ and for
magnetic fields greater than $0.5H_{c2}(T)$ the experimental data can quantitatively be
accounted for by the Coffey and Clem model, provided that the
upper-critical-field anisotropy is taken into due account.
 \PACS{
      {74.25.Ha}{Magnetic properties} \and
      {74.25.Nf}{Response to electromagnetic fields (nuclear magnetic
      resonance, surface impedance, etc.)} \and
      {74.25.Op}{Mixed states, critical fields, and surface sheaths}
     }
}
\authorrunning{M. Bonura et \textit{al.}}
\titlerunning{Microwave surface resistance in MgB$_2$}
\maketitle

\section{Introduction}

It has widely been shown that the superconducting properties of \mgb\ are strongly related to the
two-gap structure of its electronic states~\cite{sologu,Jin-lambda,bouquet-H,gurevich,golubovHc2,golubov}. The smaller
superconducting gap, $\Delta_{\pi}$, arises from the quite
isotropic $\pi$ bands and the larger one, $\Delta_{\sigma}$, from
the strongly anisotropic $\sigma$ bands~\cite{liu}. Due to
the different parity of the $\sigma$ and $\pi$ bands, inter-band
scattering of quasiparticles is much smaller than the intra-band
one, making the two superconducting gaps quite different though
they close at the same $T_c$~\cite{Mazin}.

According to the theory of multi-band
superconductivity~\cite{liu}, the inclusion of defects would increase the inter-band scattering
and, consequently, change the relative magnitude of the different
gaps. In order to carry out investigation on this topic,
essentially two methods have been used to insert defects and/or
disorder in \mgb: chemical substitution and damage by
irradiation~\cite{sostituzioni,wilke,gandikota,ghigo,tarantini1}. In any
cases, the inclusion of defects, besides to change the inter-band
scattering, might increase the upper critical field and the
critical current, strongly affecting the fluxon dynamics. Very recently, the effects of neutron
irradiation have extensively been investigated on polycrystalline
$\mathrm{Mg}^{11}
\mathrm{B}_2$~\cite{tarantini1,pallecchi,gonnelli2,putti2,puttiSUST2008}.
It has been shown that irradiation leads to an improvement in both
critical field and critical current density for an exposure level
in the range $1 \div 2\times10^{18}~\mathrm{cm}^{-2}$. On further
increasing the neutron fluence, all the superconducting
properties, such as $T_c$, $H_{c2}$, $J_c$, are strongly
suppressed. Furthermore, measurements of specific heat, as well as
point-contact spectroscopy, have shown that in the sample
irradiated at the highest fluence ($1.4 \times
10^{20}~\mathrm{cm}^{-2}$) the two gaps merge into a single
value~\cite{gonnelli2,putti2}.

Despite the large amount of experimental and theoretical work done
on \mgb, some arguments are still under discussion, such as the
effects of the magnetic field on the superconducting
properties~\cite{Magnetiz,Jia,Yang}. The main difficulty to
quantitatively discuss the mixed-state properties of \mgb\ arises
from the unusual flux-line properties due to the different
coherence lengths, $\xi_{\sigma}$ and $\xi_{\pi}$, associated with
$\Delta_{\sigma}$ and $\Delta_{\pi}$~\cite{koshelev}.
Scanning tunnelling spectroscopy on \mgb\ single crystals along
the c-axis, which probes mainly the $\pi$ band, has highlighted a
core size much larger than the estimates based on the measured
$H_{c2}$ values, as well as a significant core overlap at fields
much lower than the macroscopic $H_{c2}$~\cite{eskil}.
Furthermore, measurements of neutron scattering from the vortex
lattice have highlighted a spatial rotation of the vortex lattice
for applied magnetic fields in the range $0.5 \div
1$~T~\cite{cubit}. According to point-contact-spectroscopy
experiments~\cite{samuely,gonnelli1,diffusivita}, these unusual
properties have been ascribed to the strong suppression of the
superconductivity in the $\pi$ band, occurring in that field
range. At low fields, each vortex has a composite structure, with
$\sigma$-band quasiparticles localized in a region of radius
$\xi_{\sigma}$ and $\pi$-band quasiparticles in a wider region of
radius $\xi_{\pi}$.  In the field range $0.5 \div 1$~T (at low
$T$), the giant cores start to overlap; when the magnetic field is
large enough to suppress the $\pi$-band gap, the vortex cores
shrink and the $\pi$-band quasiparticles are widespread in the
whole sample~\cite{Jia,eskil}. This field-induced evolution of the
vortex lattice is expected to affect the vortex-vortex and
vortex-pinning interactions, making the description of the
properties involving the presence and motion of fluxons very
difficult.

The investigation of the microwave (mw) surface imped-ance, $Z_s=R_s + i
X_s$, in superconductors is a useful tool for determining several
properties of the superconducting state. In the absence of static magnetic
fields, the variation with the temperature of the condensed-fluid density
determines the temperature dependence of $Z_s$. On the other hand, the
field dependence of $R_s$ in superconductors in the mixed state is
determined by the presence of fluxons, which bring along normal fluid in
their cores, as well as the fluxon
motion~\cite{golo,TALVA,noiBKBO,CC,brandt,dulcicvecchio}. So,
investigation of the magnetic-field-induced variations of the surface
resistance provides important information on the fluxon dynamics.

Several studies of the mw response of \mgb\ reported in the literature
have shown that the experimental results cannot be accounted for
in the framework of standard
theories~\cite{Lee,emiliano,shibata,dulcic,nova,isteresiMgB2,Sarti}.
The temperature dependence of the mw conductivity, at zero DC
magnetic field, has been justified considering the coexistence of
two different superconducting fluids, one related to carriers
living on the $\sigma$ band and the other to carriers living on
the $\pi$ band~\cite{Lee,emiliano}. The magnetic field
dependence of the surface resistance has shown an anomalous
behavior especially at low temperatures; several authors have
highlighted unusually enhanced field-induced mw losses at applied
magnetic fields much lower than the upper critical
field~\cite{shibata,dulcic,nova,isteresiMgB2}. Sarti
et \emph{al}.~\cite{Sarti}, investigating the mw surface impedance
of \mgb\ film, have shown that at low fields, when the
contribution of the $\pi$-band superfluid cannot be neglected, the
magnetic-field dependence of the real and imaginary components of
the surface impedance exhibits several anomalies. Furthermore, a
magnetic hysteresis of unconventional shape has been detected in
the $R_s(H)$ curves~\cite{isteresiMgB2,EUCAS2007}. All these
results have suggested that in a wide magnetic-field range the
standard models for fluxon dynamics fail when applied to \mgb .

In this paper, we report on the microwave surface resistance of
two of the polycrystalline $\mathrm{Mg}^{11} \mathrm{B}_2$ samples
studied in Refs.~\cite{tarantini1,pallecchi,gonnelli2,putti2,puttiSUST2008}. We have investigated the unirradiated sample, which clearly shows two-gap superconductivity, and the sample irradiated at the highest neutron fluence, in which the two gaps merge into a single
value. The investigation has been carried out with the aim to
compare the results obtained in two-gap and one-gap \mgb\
superconductors. To our knowledge, the mw response of
neutron-irradiated \mgb\ samples has not yet been investigated.
The mw surface resistance has been measured as a function of the
temperature, in the range $2.5 \div 40$~K, and the DC magnetic
field, from 0 to 1~T, at increasing and decreasing values. We show
that the results obtained in the strongly irradiated sample can
quite well be justified in the framework of standard models, in
the whole ranges of temperatures and magnetic fields investigated.
On the contrary, the results obtained in the pristine sample
cannot thoroughly be justified. In particular, the $R_s(T)$
behavior at zero field has been accounted for, in the framework of
the two-fluid model, assuming a linear temperature dependence of
the normal and condensed fluid densities. At low temperatures, the
field dependence of $R_s$ has shown several anomalies, among which
a magnetic hysteresis having a unexpected shape. At temperatures
near $T_c$ and applied magnetic fields greater than $\approx
0.5H_{c2}(T)$, the results are well accounted for in the framework of
the Coffey and Clem model~\cite{CC}, with fluxons moving in the
flux-flow regime, taking into account the anisotropy of the upper
critical field.

\section{Experimental apparatus and samples}\label{sec:samples}
The microwave surface resistance, $R_s$, has been investigated in
two bulk MgB$_2$ samples. The procedure for the preparation and
irradiation of the samples is reported in detail
elsewhere~\cite{tarantini1,putti2}. The samples have been prepared by
direct synthesis from Mg (99.999\% purity) and crystalline
isotopically enriched $^{11}$B (99.95\% purity), with a residual
$^{10}$B concentration lower than 0.5\%. The use of isotopically
enriched $^{11}$B makes the penetration depth of the thermal
neutrons greater than the sample thickness, guarantying the
irradiation effect almost homogeneous over the sample. Several
superconducting properties of the samples have been reported in
Refs.~\cite{tarantini1,pallecchi,gonnelli2,putti2,puttiSUST2008}.
For simplicity and ease of comparison, we label the
two samples as in Ref.~\cite{tarantini1}, i.e.~P0 (pristine $\mathrm{Mg}^{11}
\mathrm{B}_2$) and P6 (irradiated at the highest neutron fluence).
According to point-contact spectroscopy \cite{gonnelli2} and
specific-heat measurements~\cite{putti2}, sample P0 shows a clear
two-gap superconductivity; in sample P6 the irradiation
process at very high fluence ($1.4 \times 10^{20}$ cm$^{-2}$)
determined a merging of the two gaps into a single value.

Sample P0 has a nearly parallelepiped shape with $w\approx 3.1~\mathrm{mm}$, $t\approx 1.5$~mm and $h\approx 3.2$~mm; it undergoes a narrow superconducting transition with $T_c^{onset}
\approx 39.0$~K and $\Delta T_c \approx 0.2$~K (from 90\% to 10\% of the normal-state resistivity); its residual normal-state resistivity is $\rho(40~\mathrm{K})\approx 1.6~\mu\mathrm{\Omega~cm}$ and the residual resistivity ratio RRR~$\approx$~11, the critical current density at zero magnetic
field is $J_{c0}\approx 4 \times 10^5$~{A/cm}$^2$, and $\mu_0H_{c2}(5~\mathrm{K})\approx 15~\mathrm{T}$; the anisotropy factor of the upper critical field at $T=5$~K is $\gamma \approx4.4$~\cite{tarantini1,pallecchi}.

Sample P6 has a nearly parallelepiped shape with $w\approx 1.1$~mm, $t\approx 0.8$~mm and $h\approx 1.4$~mm. The main characteristic parameters of sample P6 are: $T_c^{onset} \approx 9.1$~K, $\Delta T_c \approx 0.3$~K, RRR~$\approx$~1.1, $\rho(40~\mathrm{K})\approx 130~\mu\mathrm{\Omega~cm}$. The critical current density at $T=5$~K and at zero magnetic field is $J_{c0}\approx 3 \times 10^4$~A/cm$^2$; it exhibits a monotonic decrease with the magnetic field, following roughly an exponential law. The upper critical field is isotropic and its value at $T=5$~K is $\mu_0H_{c2}\approx 2~\mathrm{T}$.

The effects of the neutron irradiation on both superconducting and normal-state properties of a large series of Mg$^{11}\mathrm{B}_2$ bulk samples, including sample P0 and P6, have extensively been investigated in Refs.~\cite{tarantini1,pallecchi,puttiSUST2008}. On increasing the neutron fluence, it has been observed a monotonic decrease of $T_c$ and an increase of the residual normal-state resistivity $\rho(T_c)$. Nevertheless, it has been shown that the irradiation does not affect the variation of the normal-state resistivity $\Delta \rho = \rho (300~\mathrm{K})-\rho (T_c)$. As suggested by Rowell~\cite{Rowell}, just $\Delta \rho$ is a parameter that gives information on the grain connectivity.  The results reported in Ref.~\cite{tarantini1} show that $\Delta \rho$  remains of the order of $10~ \mu \mathrm{\Omega}$~cm over the whole range of irradiation level; in particular, in sample P0 $\Delta \rho =16~ \mu \mathrm{\Omega}$~cm and in sample P6  $\Delta \rho =12~ \mu \mathrm{\Omega}$~cm, indicating that thermal-neutron irradiation does not affect the grain-boundary properties. On the other hand, it has been shown that either neutron irradiation or He-ion irradiation~\cite{gandikota} do not affect the grain connectivity, even at high irradiation levels, contrary to what occurs using heavy-ion irradiation~\cite{ghigo}. Recent studies by transmission electron microscopy have highlighted that neutron irradiation in these samples creates nanometric amorphous regions (mean diameter $\sim 4$~nm) in the crystal lattice, whose density scales with the neutron dose~\cite{puttiSUST2008}. Studies on the field dependence of the critical current density have shown that at moderate neutron-fluence levels ($\leq10^{19}~\mathrm{cm}^{-2}$) such defects introduce new pinning centers, leading to an improvement of the critical current density; on the contrary, for neutron fluence higher than $10^{19}~\mathrm{cm}^{-2}$ (as for sample P6) these nanometric defects do not act as pinning centers because they are smaller than the coherence length~\cite{puttiSUST2008}. Moreover, these studies have shown that in the pristine and the heavily irradiated samples the pinning mechanism is ruled by grain boundaries. The defects induced by neutron irradiation act as inter- and intra-band scattering centers; the intra-band scattering causes a reduction of the electron mean free path and is responsible for the growth of the normal-state resistivity. The reduction of $T_c$ has been ascribed to both the scattering processes and the smearing of the electron density of states near the Fermi surface~\cite{gandikota,tarantini1}.

Although in the two samples $\Delta T_c$ is roughly the same, in sample P6, due to the reduced $T_c$ value, $\Delta T_c/T_c \approx 0.03$, affecting noticeably the temperature dependence of the mw surface resistance near $T_c$. On the other hand, from AC susceptibility measurements at 100 kHz, we have found that the first derivative of the real part of the AC susceptibility can be described by a Gaussian distribution function of $T_c$, centered at $T_{c0}=8.5\pm 0.1$~K with $\sigma_{T_c}=0.2\pm 0.05$~K. In the following, we will use this distribution function to quantitatively discuss the results obtained in sample P6. On the contrary, for sample P0, $\Delta T_c/T_c$ is one order of magnitude smaller, not noticeably affecting the $R_s(T)$ curve.

\begin{figure}[hb]
\centering \includegraphics[width=7 cm]{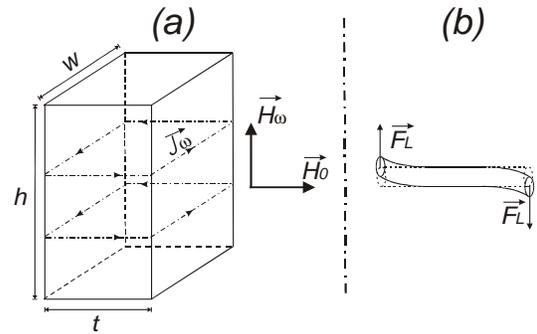} \caption{(a)
Field and current geometry at the sample surface. (b) Schematic
representation of the motion of a flux line.} \label{sample}
\end{figure}

The mw surface resistance has been measured using the
cavity-perturbation technique~\cite{golo}. A copper cavity, of
cylindrical shape with golden-plated walls, is tuned in the
$\mathrm{TE}_{011}$ mode resonating at $\omega/2\pi \approx
9.6$~GHz. The sample is located in the center of the cavity, by a
sapphire rod, where the mw magnetic field is maximum. The cavity
is placed between the poles of an electromagnet which generates DC
magnetic fields up to $\mu_0H_0\approx 1$~T. Two additional coils,
independently fed, allow compensating the residual field and
working at low magnetic fields. A liquid-helium cryostat and a
temperature controller allow working either at fixed temperatures
or at temperature varying with a constant rate. The sample and the
field geometries are shown in Fig.~1a; the DC magnetic field is
perpendicular to the mw magnetic field,
$\emph{\textbf{H}}_{\omega}$. When the sample is in the mixed
state, the induced mw current causes a tilt motion of the whole vortex
lattice~\cite{brandt}; Fig.~\ref{sample}b schematically shows the
motion of a flux line.

The surface resistance of the sample is given by\\
\begin{equation*}
    R_s= \Gamma~\left(\frac{1}{Q_L} - \frac{1}{Q_U}\right)\,,
\end{equation*}
where $Q_L$ is the quality factor of the cavity loaded with the
sample, $Q_U$ that of the empty cavity and $\Gamma$ the geometry
factor of the sample.\\ The quality factor of the cavity has been
measured by an hp-8719D Network Analyzer. The surface resistance
has been measured as a function of the temperature, at fixed
values of the DC magnetic field, and as a function of the field,
at fixed temperatures. All the measurements have been performed at very low input power; the estimated amplitude of the mw magnetic field in the region in which the sample is located is of the order of $0.1~\mu$T.

\section{Experimental results}\label{experimental}
\begin{figure}[ht]
\centering \includegraphics[width=8 cm]{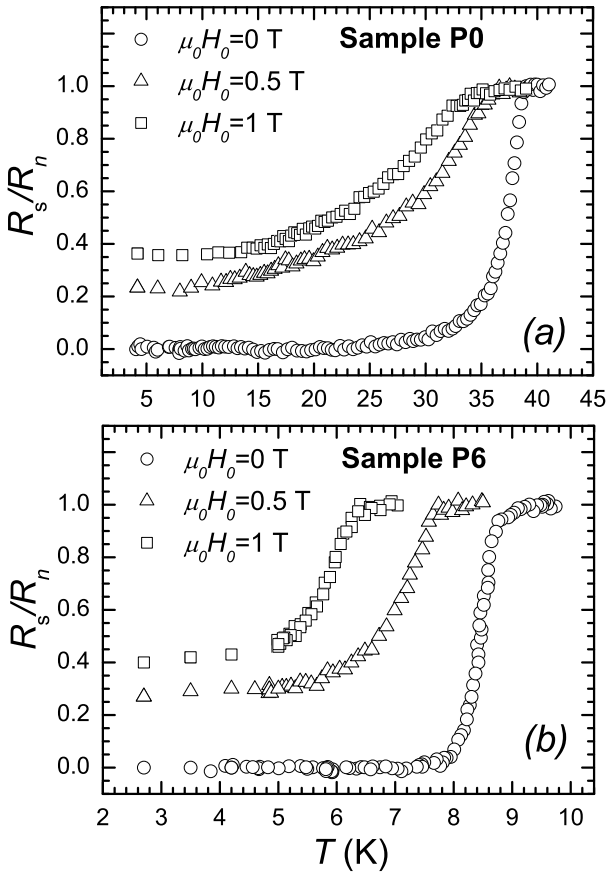} \caption{Normalized
values of the surface resistance as a function of the temperature,
obtained in the two samples, at different values of the DC field. $R_n$ is
the surface resistance at $T = T_c$.} \label{RS(T)}
\end{figure}

Figure~\ref{RS(T)} shows the temperature dependence of the surface
resistance in the pristine (a) and irradiated (b) \mgb\ samples,
at different values of the DC magnetic field. In order to
disregard the geometry factor, and compare the results in samples
of different dimensions, we have normalized the data to the value
of the surface resistance in the normal state, $R_n$, at
$T=T_c^{onset}$. The results have been obtained according to the
following procedure: the sample was zero-field cooled (ZFC) down
to low temperature, then $H_0$ was set at a given value and kept
constant during the time the measurement has been performed.

On increasing $H_0$, the $R_s(T)$ curves broaden and shift towards
lower temperatures; however, the effects of the applied magnetic
field is different in the two samples. Although the value of
$H_{c2}$ of sample P0 at low temperatures is one order of
magnitude larger than that of P6, the field-induced variations of
$R_s$ in the two samples have roughly the same magnitude. In
sample P0, one can observe an anomalously enhanced field-induced
broadening of the $R_s(T)$ curve, which extends down to the lowest
temperature. On the contrary, in sample P6 the larger shift of the
$R_s(T)$ curve induced by $H_0$ is expected because of the lower
$H_{c2}$ value.

\begin{figure}[t]
\centering
\includegraphics[width=8cm]{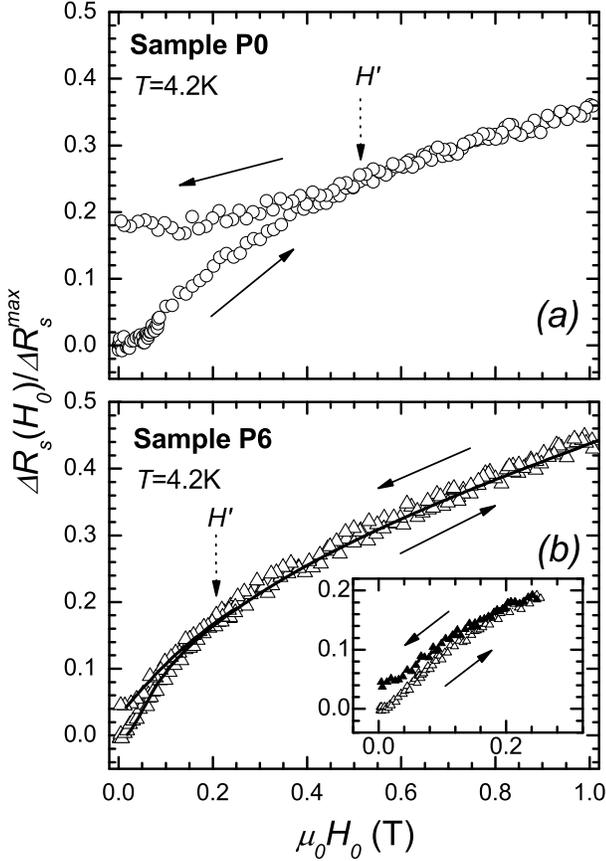}
\caption{Field-induced variations of $R_s$ for samples P0 (a) and
P6 (b), at $T=4.2$~K. $\Delta R_s(H_0)\equiv R_s(H_0,T)-R_{res}$,
where $R_{res}$ is the residual mw surface resistance at $T=2.5$~K
and $H_{0}=0$; $\Delta R_s^{max}\equiv R_{n}-R_{res}$. The line is
the best-fit curve obtained, as explained in
Sec.~\ref{discussion}, with $\mu_0 H_{c2}= 1.71$~T,
$\omega_0/\omega = 0.67$ and the field dependence of the critical
current density reported in Ref.~\cite{tarantini1}. The inset shows a minor
hysteresis loop obtained by sweeping $H_0$ from 0 to 0.25~T and
back.} \label{T4K}
\end{figure}

\begin{figure}
\centering
\includegraphics[width=8cm]{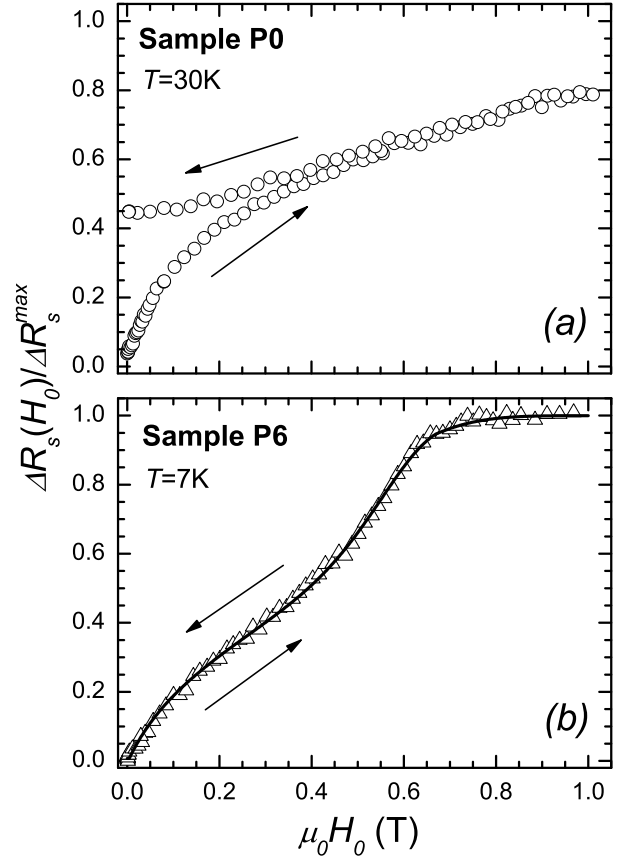}
\caption{Normalized field-induced variations of $R_s$ for samples
P0 (a) and P6 (b), at $T/T_c \approx 0.77$. The line in panel (b)
is the best-fit curve of the data, obtained as described in the
text by using the field dependence of the depinning frequency
reported in Fig.~\ref{frequenza-depinning}.} \label{Tintermed}
\end{figure}

\begin{figure}
\centering
\includegraphics[width=8cm]{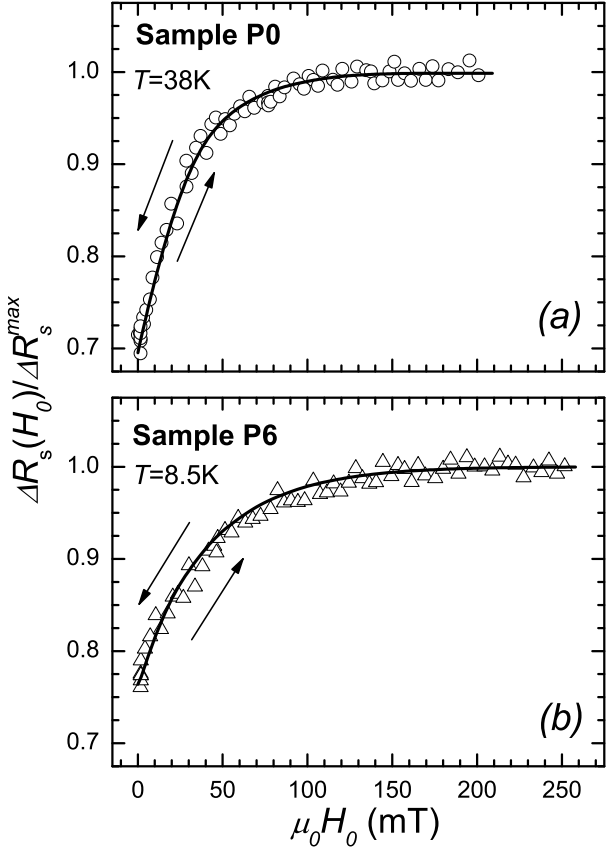}
\caption{Normalized field-induced variations of $R_s$ for samples
P0 (a) and P6 (b), at temperatures close to $T_c$. Lines are
best-fit curves, obtained by the model of Sec.~\ref{model}
considering fluxons move in the flux-flow regime. The line of
panel (a) has been obtained taking into account the anisotropy of
the upper critical field, as described in the text, with
$\gamma=3.3$; for sample P0, we have set $\gamma=1$ consistently
with the results of Ref.~\cite{pallecchi}.} \label{TnearTc}
\end{figure}

The field-induced variations of $R_s$ have been investigated for
different values of the temperature. For each measurement, the
sample was ZFC down to the desired temperature; the DC magnetic
field was increased up to a certain value and, successively,
decreased down to zero. Figures~\ref{T4K}, \ref{Tintermed} and
\ref{TnearTc} show the field-induced variations of $R_s$ for the
two samples, at different temperatures. In all the figures,
$\Delta R_s(H_0)\equiv R_s(H_0,T)-R_{res}$, where $R_{res}$ is the
residual mw surface resistance at $T=2.5$~K and $H_{0}=0$;
moreover, the data are normalized to the maximum variation,
$\Delta R_s^{max}\equiv R_{n}-R_{res}$. The continuous lines
reported in the figures are the best-fit curves obtained by the
model reported in Sec.~\ref{model}.

In both samples, $R_s$ does not show any variation as long as the
magnetic field reaches a certain value, depending on $T$, that
identifies the first-penetration field, $H_p$. For $H_0>H_p$,
vortices start to penetrate the sample and, consequently, $R_s$
increases.

Figure~\ref{T4K} refers to the results obtained at $T=4.2$~K. At
this temperature, in both samples the $R_s(H_0)$ curves exhibit a
magnetic hysteresis, which disappears for $H_0$ higher than a
certain value, indicated in the figure as $H ^{\prime}$. The inset
in panel (b) shows a minor hysteresis loop obtained by sweeping
$H_0$ from 0 to 0.25~T and back. The field-induced variations of
$R_s$ in sample P0 show some anomalies. Firstly, the application
of a magnetic field of $\approx 1$~T, which is about $H_{c2}/15$,
causes a $R_s$ variation of $\approx 35\%$ of the maximum
variation. These field-induced variations of $R_s$ are much
greater than those expected from the models reported in the
literature~\cite{CC,brandt,dulcicvecchio} and detected in other
superconductors~\cite{TALVA,noiBKBO}. A comparison with the
results of panel (b) shows that in sample P6 a $R_s$ variation of
the same order is obtained for the same value of $H_0$, even
though, in this case, 1 T is about $H_{c2}/2$. Results similar to
those obtained in P0 have been observed in other \mgb\ samples,
produced by different methods and, therefore, seems to be a
peculiarity of
\mgb~\cite{shibata,nova,isteresiMgB2,EUCAS2007}. The
finding that in sample P6 we have not observed this anomalous
result strongly suggests that the enhanced $R_s$ variation is due
to the two-gap superconductivity.

A magnetic hysteresis of $R_s$ is expected in superconducting
samples in the critical state; it is ascribable to the different
magnetic induction at increasing and decreasing DC
fields~\cite{noiisteresi}. Most likely, the different
amplitude of the hysteresis loop obtained in the two samples is
due to the different values of the critical current density; a
smaller hysteresis is observed in sample P6 because of the smaller
$J_c$ value. However, as it is visible in Fig.~\ref{T4K}, also the
shape of the hysteresis loop is different in the two samples. The
decreasing-field branch of the $R_s(H_0)$ curve in sample P6 has a
negative concavity down to $H_p$, as expected~\cite{noiisteresi}.
On the contrary, in sample P0 one can observe a plateau, in the
field range $0\div 0.2$~T, which cannot be justified in the
framework of the critical-state models, considering the measured
field dependence of $J_c$~\cite{tarantini1}. We would like to remark that
this result has been obtained in all of the \mgb\ samples we have
investigated~\cite{isteresiMgB2,EUCAS2007}.

Figure~\ref{Tintermed} shows the field-induced variations of
$R_s$, for sample P0 (a), at $T=30$~K, and for sample P6 (b), at
$T=7$~K; for both samples, $T/T_c \approx 0.77$. In the $R_s(H_0)$
curve of sample P0 the hysteresis is still present, probably due
to the high value of $J_c$ at this temperature, and still has an
anomalous shape. We remark that in sample P0 we have
observed magnetic hysteresis of $R_s$ up to $T/T_c \approx 0.95$,
while in sample P6 the hysteresis becomes undetectable at $T/T_c
\gtrsim 0.55$.

Figure~\ref{TnearTc} shows the field-induced variations of $R_s$
at temperatures near $T_c$, where in both samples the $R_s(H_0)$
curve is reversible. As will be shown in Sec.~\ref{discussion}, at
temperatures near $T_c$ the field dependence of the mw surface
resistance can be accounted for by standard models also for sample
P0, provided that the anisotropy of the upper critical field is
taken into account.

\begin{figure}
\includegraphics[width=8cm]{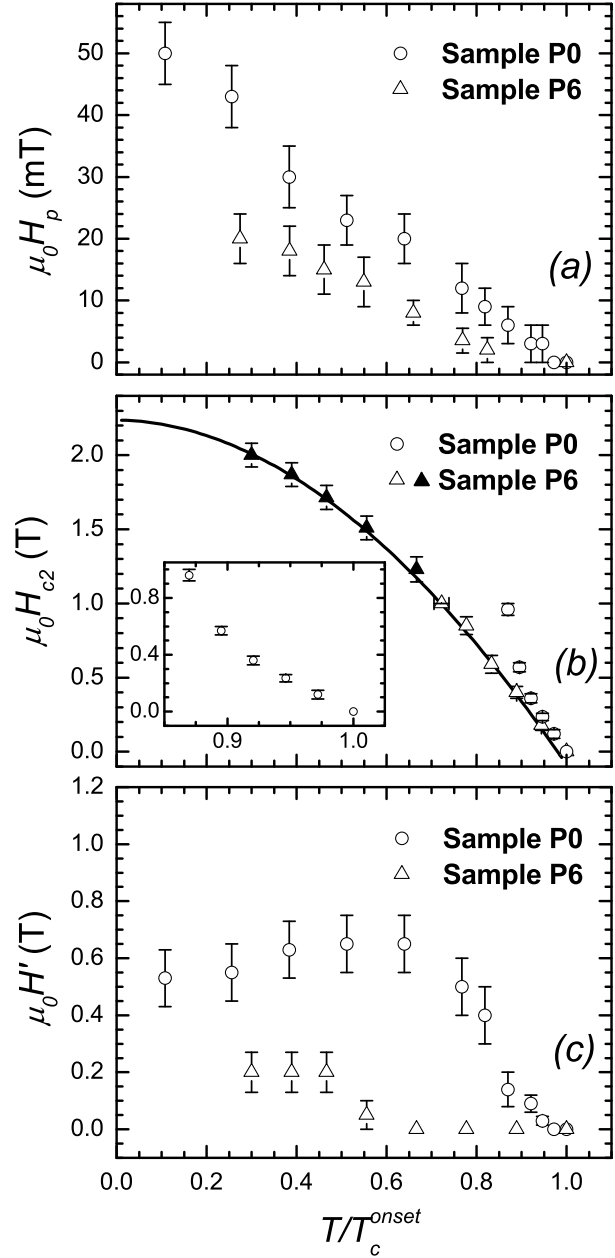}
\caption{Temperature dependence of the characteristic fields,
$H_p$, $H_{c2}$ and $H^{\prime}$, for the two samples.  In panel
(b): the inset shows the $H_{c2}(T)$ values of sample P6 in an
enlarged scale; the continuous line is the best-fit curve of the
experimental data, obtained for sample P6, as described in the
text.} \label{campicaratteristici}
\end{figure}

From isothermal $R_s(H_0)$ curves, obtained at different
temperatures, we have deduced the temperature dependence of the
characteristic fields, $H_p$, $H_{c2}$ and $H^{\prime}$. In
Fig.~\ref{campicaratteristici} we report the values of $H_p$,
$H_{c2}$ and $H^{\prime}$ as a function of the reduced
temperature, $T/T_c^{onset}$, for the two samples. The inset in
panel (b) shows $H_{c2}(T)$ of sample P6 in an enlarged scale.

From Fig.~\ref{campicaratteristici}a, one can see that $H_p$ of
sample P0 exhibits a linear temperature dependence down to low
temperatures, consistently with results reported by different
authors in bulk~\cite{Li,sharoni} and
crystalline~\cite{caplin,Lyard} \mgb\ samples. The extrapolated
value at $T=0$ is about 55~mT; so, considering the demagnetization
effect, the estimated value of the lower critical field is
$H_{c1}(0)\approx70$~mT. This value, although consistent with the
lower critical field reported for \mgb\ crystals by some
authors~\cite{Lyard}, is slightly larger than that reported for
bulk samples, which ranges from 15 to 45~mT. We suggest that this
is ascribable to weak surface-barrier effects.

In sample P6 we have obtained $H_p$ values smaller, but of the same order, than those of
sample P0; this may be due to the irradiation effects. Indeed, the
authors of Ref.~\cite{Magnetiz}, from magnetization measurements in
neutron-irradiated \mgb\ crystals, have observed that the lower
critical field reduces monotonically on increasing the fluence.

The values of $H_{c2}(T)$ indicated in
Fig.~\ref{campicaratteristici}b as open triangles have been
deduced measuring the magnetic field at which $R_s$ reaches the
normal state value, $R_n$. At the temperatures in which the upper
critical field of sample P6 is higher than the maximum magnetic
field achievable with our experimental apparatus ($\approx 1$~T),
the $H_{c2}(T)$ values (full triangles in the figure) have been
obtained as best-fit parameters, using the model reported in
Sec.~\ref{model}. On the contrary, since the results obtained in
sample P0 cannot be accounted for by the model (except at
temperature close to $T_c$), we report only the values we have
directly deduced from the experimental results. For $T \ge 5$~K,
the values we obtained for $H_{c2}(T)$, in both samples, agree with those reported
in Ref.~\cite{tarantini1} (the authors do not report $H_{c2}$ at lower
temperatures). Our results give complementary information about
the temperature dependence of $H_{c2}$ of sample P6 at low
temperatures. The continuous line in
Fig.~\ref{campicaratteristici}b has been obtained by fitting the
data of sample P6 with $H_{c2}(T)=H_{c20}[1-(T/T_c)^{\alpha}]$; we
have obtained, as best-fit parameters, $\mu_0
H_{c20}=(2.2\pm0.2)$~T, $\alpha=1.9\pm0.3$ and
$T_c=(8.9\pm0.2)$~K. This temperature dependence of the upper
critical field is consistent with that expected in conventional
superconductors. On the contrary, for sample P0 we observed an
upward curvature of $H_{c2}(T)$, clearly visible in the inset,
characteristic of two-gap \mgb\
materials~\cite{sologu,gurevich,golubovHc2}.

$H^{\prime}(T)$ of Fig.~\ref{campicaratteristici}c corresponds to
the value of the DC magnetic field at which the decreasing-field
branch of the $R_s(H_0)$ curves deviates from the increasing-field
branch. The zero values (without error bar) mean that the
hysteresis is not detectable at the corresponding temperatures.
Consistently with the lower value of the critical current density,
$H^{\prime}(T)$ is smaller in sample P6 than in P0. We would like to
remark that the values of $H^{\prime}(T)$ could differ from the
irreversibility field deduced from magnetization measurements.
Indeed, it has been shown that, in samples of finite dimensions,
the application of an AC magnetic field normal to the DC field can
induce the fluxon lattice to relax toward an uniform
distribution~\cite{brandt3}. Furthermore, measurements we have
performed in different superconducting samples have pointed out
that, for samples of millimetric size, the sensitivity of our
experimental apparatus does not allow resolving magnetic
hysteresis of $R_s$ when $J_c < 10^{4}~\mathrm{A/cm}^2$. By
considering the values of $J_c$ reported in Ref.~\cite{tarantini1}, at $T=4.2$~K
we should obtain $\mu_0 H^{\prime}\approx 0.2$~T for sample P6 and
$\mu_0 H^{\prime} \approx 3.5$~T for sample P0. From
Fig.~\ref{campicaratteristici}c, one can see that this expectation
is verified in sample P6; on the contrary, the value of
$H^{\prime}$ in sample P0 is about one order of magnitude smaller
than the expected one.

\section{The model} \label{model}
Microwave losses induced by static magnetic fields have been
investigated by several
authors~\cite{golo,TALVA,noiBKBO,CC,brandt,dulcicvecchio,noiisteresi,noistatocritico}.
At low temperatures and for applied magnetic fields lower enough
than the upper critical field, the main contribution arises from
the fluxon motion; however, it has been pointed out that a
noticeable contribution can arise from the presence of normal
fluid, especially at temperatures near $T_c$ and for magnetic
fields of the same order of $H_{c2}(T)$. The majority of the
models assume an uniform distribution of fluxons inside the
sample; so, they disregard the effects of the critical state. Very
recently, we have investigated the field-induced variations of the
mw surface resistance in superconductors in the critical
state~\cite{noiisteresi,noistatocritico}, and have accounted for
the magnetic hysteresis in the $R_s(H_0)$ curves.

In the London local limit, the surface resistance is proportional
to the imaginary part of the complex penetration depth,
$\widetilde{\lambda}$, of the em field:\\
\begin{equation}\label{Rs}
    R_s=-\mu_{0}\omega ~\mathrm{Im}[{\widetilde{\lambda}(\omega,B,T)}].
\end{equation}
The complex penetration depth has been calculated in different
approximations~\cite{CC,brandt}. Coffey and Clem (CC) have
elaborated a comprehensive theory for the electromagnetic response
of superconductors in the mixed state, in the framework of the
two-fluid model of superconductivity~\cite{CC}. The theory has
been developed under two basic assumptions: i) inter-vortex
spacing much less than the field penetration depth; ii) uniform
vortex distribution in the sample. With these assumptions vortices
generate a magnetic induction field, $B$, uniform in the sample.
This approximation is valid for $H_0 > 2H_{c1}$ whenever the
fluxon distribution can be considered uniform within the AC
penetration depth.

In the linear approximation, $H_{\omega} \ll\ H_0$,
$\widetilde{\lambda}(\omega,B,T)$ expected from the CC model is
given by
\begin{equation}\label{lambdat}
    \widetilde{\lambda}(\omega,B,T)=\sqrt{\frac{\lambda^{2}(B,T)+
    (i/2)\widetilde{\delta}_{v}^{2}(\omega,B,T)}
    {1-2i\lambda^{2}(B,T)/\delta _{nf}^{2}(\omega,B,T)}}\, ,
\end{equation}
with
\begin{equation}\label{lamda0}
\lambda(B,T) = \frac{\lambda_0}{\sqrt{[1-w_0(T)][1- B
/B_{c2}(T)]}}\,,
\end{equation}
\begin{equation}\label{delta0}
\delta_{nf}(\omega,B,T) =
\frac{\delta_0(\omega)}{\sqrt{1-[1-w_0(T)][1- B /B_{c2}(T)]}}\,,
\end{equation}
where $\lambda_0$ is the London penetration depth at  $T = 0$,
$\delta_0$ is the normal-fluid skin depth at $T = T_c$, $w_0(T)$
is the fraction of normal electrons at $H_0 = 0$; in
the Gorter and Casimir two-fluid model $w_0(T)=(T/T_c)^4$.\\
$\widetilde{\delta} _{v}$ is the effective complex skin depth
arising from the vortex motion; it depends on the relative
magnitude of the viscous and restoring-pinning forces, which identities the depinning
frequency $\omega_0$.
$\widetilde{\delta} _{v}$ can be written as\\
\begin{equation}\label{delta-v(omega)}
\frac{1}{\widetilde{\delta}_{v}^{2}}=\frac{1}{\delta_{f}^{2}}\left(1+
i~\frac{\omega_0}{\omega}\right)\,,
\end{equation}
where
\begin{equation}\label{delta-f}
\delta_{f}^{2}=\frac{2B\phi_0}{\mu_0 \omega \eta}\,,
\end{equation}
with $\eta$ the viscous-drag coefficient and $\phi_0$ the quantum of flux.\\
When the frequency of the em wave, $\omega$, is much lower than
$\omega_0$, the fluxon motion is ruled by the restoring-pinning
force. On the contrary, for $\omega \gg \omega_0$, the fluxon
motion takes place around the minimum of the pinning-potential
well and, consequently, the restoring-pinning force is nearly
ineffective. So, the contribution of the viscous-drag force
predominates and the induced em current makes fluxons move in the
flux-flow regime. In this case, enhanced field-induced energy
losses are expected.

As it is clear from Eqs.~(1--4), it is expected that the features
of the $R_s(H_0)$ curves strongly depend on the applied-field
dependence of $B$. On the other hand, the CC theory is strictly
valid when $B$ is uniform inside the sample; in particular for
$H_0 \gg H_{c1}$, the $R_s(H_0)$ curves can be described setting
$B=\mu_0 H_0$. When fluxons are in the critical state, the
assumption of uniform $B$ is no longer valid and the CC theory
does not correctly describe the field-induced variations of $R_s$.
As a consequence, the hysteresis in the $R_s(H_0)$ curve cannot be
justified by Eqs.~(1--4). In our field geometry (see
Fig.~\ref{sample}a), the effects of the non-uniform $B$
distribution on $R_s$ are particularly enhanced because in the two
surfaces of the sample normal to the external magnetic field the
mw current and fields penetrate along the fluxon axis and,
consequently, the mw losses involve the whole vortex lattice.
However, in this case, one can easily take into account the
non-uniform $B$ distribution by calculating a proper averaged
value of $R_s$ over the whole sample as
follows~\cite{noiisteresi,noistatocritico}
\begin{equation}\label{RsMED}
    R_{s}= \frac{1}{S}\int_\Sigma R_s(|B(\textit{\textbf{r}})|)\,
    dS\,,
\end{equation}
where $\Sigma$ is the sample surface, $S$ is its area and
\textit{\textbf{r}} identifies the surface element.\\
The pinning effects are particularly enhanced
at temperatures smaller enough than $T_c$, where the dissipations
are essentially due to vortex motion. So, the main contribution to
$R_s$ comes from the sample regions in which fluxons experience
the Lorentz force due to the mw current, i.e. where
$\textit{\textbf{H}}_0 \times \textbf{\textit{J}}_\omega \neq 0$.
Furthermore, in order to take into due account the critical-state
effects by Eq.~(\ref{RsMED}), it is essential to know the $B$
profile inside the sample, determined by $J_c(B)$.

Recently, using this method, we have investigated the effects of
the critical state on the field-induced variation of $R_s$, at
increasing and decreasing
fields~\cite{noiisteresi,noistatocritico}. We have shown that the
parameter that mainly determines the peculiarities of the
$R_s(H_0)$ curve is the full penetration field, $H^*$. Firstly,
the width of the hysteresis is directly related to the value of
$H^*$; samples of small size and/or small $J_c$ are expected to
exhibit weak hysteretic behavior. Furthermore, $H^*$ determines
the shape of the hysteresis loop as well. On increasing the
external field from zero up to $H^*$, more and more sample regions
contribute to the mw losses; this gives rise to a positive
curvature of the increasing-field branch of the $R_s(H_0)$ curve.
For $H_0 > H^*$, in the whole sample the local magnetic induction
depends about linearly on the external field and the
increasing-field branch is expected to have a negative concavity.
The shape of the decreasing-field branch is strictly related to
the shape of the magnetization curve; it should exhibit a negative
concavity, with a monotonic reduction of $R_s$ in the whole field
range swept.

\section{Discussion}\label{discussion}
As we have shown in Sec.~\ref{experimental}, the $R_s(H_0, T)$
curves exhibit different peculiarities in the unirradiated sample
(P0) and the strongly irradiated sample (P6). The model described
in Sec.~\ref{model} fully justifies the experimental results
obtained in sample P6, which exhibits a single-gap
superconductivity~\cite{gonnelli2,putti2}. On the contrary, the
results obtained in sample P0 cannot be justified in the framework
of the same model, either for the magnetic-field dependence or 
for the temperature dependence of the surface resistance, even at
zero DC field. Only the results obtained at temperatures close to
$T_c$ can be justified, provided that the anisotropy of the upper
critical field is taken into due account. In the following,
firstly we will discuss the temperature dependence of the mw
surface resistance in the absence of DC magnetic fields;
successively, we will discuss the field-induced variations of
$R_s$.

\subsection{Temperature dependence of $R_s$ in zero magnetic
field} 
\begin{figure}
  \includegraphics[width=8cm]{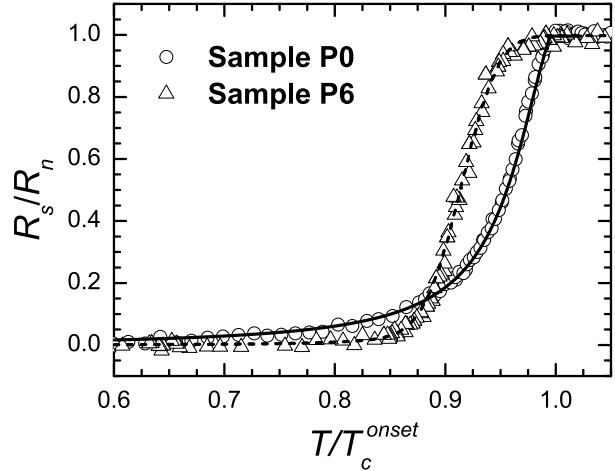}\\
  \caption{Normalized values of the mw surface resistance as a function of the reduced
  temperature, obtained in the two samples at $H_0=0$. Symbols are experimental data;
  lines are the best-fit curves obtained as described in the text.}\label{RST-FIT}
\end{figure}

Figure~\ref{RST-FIT} shows the normalized values of the
surface resistance at $H_0=0$ as a function of the reduced
temperature for both samples. The $R_s(T)$ curve of sample P6
shows a wide transition, broadened in a roughly symmetric way with
respect to the middle point at $R_s/R_n = 0.5$. This behavior can
be ascribed to the $T_c$ distribution over the sample. On the
contrary, in sample P0 one can notice a sharp variation of
$R_s(T)$, at temperatures near $T_c$, and a wide tail, extending
from $T/T_c \approx 0.9$ down to $T/T_c \approx 0.7$, which cannot
be ascribed to the $T_c$ distribution. The lines in the figure are
best-fit curves; they have been obtained with different procedures
for the two samples.

The results obtained from the model discussed in Sec.~\ref{model}
setting $B=0$ in Eqs.~(\ref{Rs}--\ref{delta0}) converge to those
of the em response of superconductors in the Meissner state, in
the framework of the two-fluid model. In this case, the
temperature dependence of $R_s/R_n$ is determined, apart from the
$T_c$ distribution over the sample, by the temperature dependence
of the normal-fluid density, $w_0(T)$, and the ratio $\lambda_0 /
\delta_0$. In order to fit the experimental data obtained in
sample P6, we have assumed $w_0(T) = (T/T_c)^4$, consistently with
the Gorter and Casimir two-fluid model, and have used
Eqs.~(\ref{Rs}--\ref{delta0}) with $B=0$. We have averaged the
expected curve over a gaussian distribution function of $T_c$ with
$T_{c0}=8.5$~K and $\sigma=0.2$~K (see Sec.~\ref{sec:samples}) and
have used $\lambda_0 / \delta_0$ as fitting parameter. The
best-fit curve, dashed line in Fig.~\ref{RST-FIT}, has been
obtained with $\lambda_0 / \delta_0=0.14$; however, we have found
that the expected curve is little sensitive to variations of
$\lambda_0 / \delta_0$, except at low temperatures, where the
measured $R_s$ is limited by the sensitivity of our experimental
apparatus. In particular, by varying $T_{c0}$ and $\sigma$ within
the experimental uncertainty, good agreement is obtained with
$\lambda_0 / \delta_0$ values ranging from 0.04 to 0.15. This
occurs because the $T_c$ distribution broadens the $R_s(T)$ curve,
hiding the $\lambda_0 / \delta_0$ effects.

Unlike for sample P6, the results of Fig.~\ref{RST-FIT} obtained
in P0 cannot be justified in the framework of the Gorter and
Casimir two-fluid model, using reasonable values of $\lambda_0 /
\delta_0$. On the other hand, different authors
\cite{Jin-lambda,golubov,moca} have shown that the temperature
dependence of the field penetration depth in MgB$_2$ cannot be
accounted for by either the Gorter and Casimir two-fluid model or
the standard BCS theory. A linear temperature dependence of the
condensed fluid density, in a wide range of temperatures below
$T_c$, has been reported, which has been justified in the
framework of two-gap models for the MgB$_2$
superconductor~\cite{golubov,moca}. Prompted by these
considerations, we have hypothesized a linear temperature
dependence of $w_0$. The continuous line in Fig.~\ref{RST-FIT} is
the best-fit curve; it has been obtained by
Eqs.~(\ref{Rs}--\ref{delta0}) with $B=0$, $w_0(T)=T/T_c$ and
$\lambda_0 / \delta_0=0.15$. The wide low-$T$ tail is essentially
determined by the linear temperature dependence of $w_0$. The
sensitivity achievable by our experimental apparatus does not
allow determining the small variations of $R_s(T)$ for $T/T_c
\lesssim 0.5$; so, from $R_s(T)$ measurements no indication about
the temperature dependence of the densities of the normal and
condensed fluids at low temperatures can be obtained. However, the
linear temperature dependence of the lower critical field we
obtained (see Fig.~\ref{campicaratteristici}a) strongly suggests
that $w_0$ linearly depends on $T$ down to low temperatures.

\subsection{Field dependence of $R_s$ in sample P6}
In conventional (single-gap) superconductors, it is expected that the field
dependence of the mw surface impedance is described by the model
reported in Sec.~\ref{model}. In this framework, in order to
calculate the expected field-induced variations, by
Eqs.~(\ref{Rs}--\ref{delta0}), the essential parameters are the
value of $\lambda_0 / \delta_0$, $H_{c2}(T)$, the depinning
frequency, $\omega_0$, and its field dependence. It is not
necessary to consider the upper-critical-field anisotropy,
$\gamma$, because it has been shown that in the P6 sample
$\gamma=1$~\cite{pallecchi}. When the critical-state effects
cannot be neglected, in order to use Eq.~(\ref{RsMED}), it is also
essential to know the profile of the induction field determined by
the field dependence of the critical current density. The value of
$\lambda_0 / \delta_0$ has been determined by fitting the $R_s(T)$
curve at $H_0=0$; the critical current density and its field
dependence are reported in Ref.~\cite{tarantini1}. The values of $H_{c2}(T)$ at
$T\geq5$~K are reported in Ref.~\cite{tarantini1}, and/or deduced from our
experimental data; at $T<5$~K, $H_{c2}$ has to be considered as
parameter. It is worth noting that the large uncertainty of
$\lambda_0 / \delta_0$, we obtained for this sample, does not
affect the best-fit curves because this parameter essentially
determines the normalized $\Delta R_s(H_0)$ value at $B=0$.

\begin{figure}
  \includegraphics[width=8cm]{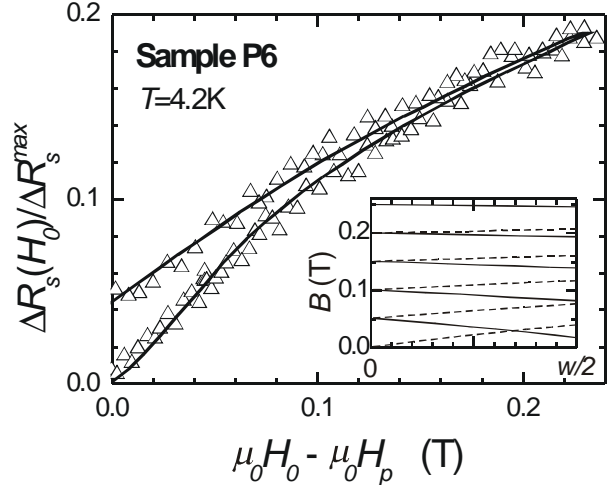}
  \caption{Field-induced variations of the mw resistance, obtained
  in sample P6 by sweeping the magnetic field from 0 to 0.25~T and back.
  The line is the best-fit curve obtained, as explained in the text, with
  $\mu_0 H_{c2}= 1.71$~T, $\omega_0/\omega = 0.67$ and the field dependence
  of the critical current density reported in Ref.~\cite{tarantini1}. The inset
  shows the $B$ profile at increasing (---) and decreasing (-~-) fields; $w$ is the
  width of the sample.}
  \label{minorloop}
\end{figure}

As one can see from Fig.~\ref{T4K}, at $T=4.2$~K the $R_s(H_0)$
curve exhibits a magnetic hysteresis, indicating that the effects
of the critical state are not negligible. In order to use
Eq.~(\ref{RsMED}), we have calculated the $B$ profile in the
sample using the field dependence of the critical current,
$J_c(B)$, reported in Ref.~\cite{tarantini1} and we have set the induction field
at the edges of the sample as $B= \mu_0 (H_0-H_p)$; we have taken
$H_{c2}$ and $\omega_0$ as fitting parameters. The line of
Fig.~\ref{T4K}b is the best-fit curve; it has been obtained with
$\mu_0 H_{c2}= 1.71$~T and $\omega_0/\omega = 0.67$ independent of
$H_0$. For the sake of clarity, in Fig.~\ref{minorloop} we report
the results obtained by sweeping the DC magnetic field from 0 to
0.25~T and back, along with the expected curve. The inset shows
the $B$ profile along the width of the sample at half height,
determined by $J_c(B)$; the continuous lines are the
increasing-field profiles, the dashed ones are the
decreasing-field profiles at the same external-field values. As
one can see, taking into account the field distribution inside the
sample, the experimental results are quite well justified in the
framework of the model discussed in Sec.~\ref{model}. In the increasing-field branch, a change of concavity is well visible at $\mu_0(H_0-H_p)\approx 0.04$~T, consistently with the
expected value of the full penetration field; the decreasing-field
branch exhibits a negative concavity in the whole range of fields.

Following the procedure above described, from the best fit of the
experimental data of the isothermal $R_s(H_0)$ curves at $T < 6$~K
we have obtained the $H_{c2}(T)$ values indicated as full
triangles in Fig.~\ref{campicaratteristici}b.

When the $R_s(H_0)$ curves do not show hysteresis, the effects of
the critical state are negligible and the induction field, $B$,
can be considered uniform. In this case,
we have used the following approximate expression for the magnetization:\\
\begin{equation*}\label{m}
    M=-H_p+ \frac{H_p}{H_{c2}-H_p}(H_0-H_p)\,;
\end{equation*}
and, consequently
\begin{equation*}\label{B}
    B=\mu_0 \left(1+\frac{H_p}{H_{c2}-H_p}\right)(H_0-H_p).
\end{equation*}

Several calculations have shown that, in order to fit the
experimental data at $T \geq7 $~K, it is essential to consider the
$T_c$ distribution over the sample. So, we have averaged the
expected $R_s(H_0)$ curves [calculated by
Eqs.~(\ref{Rs}--\ref{delta0})] over the $T_c$ distribution (see
Sec.~\ref{sec:samples}). We have used for $H_p(T)$ and $H_{c2}(T)$
the values deduced from the experimental data, letting them vary
within the experimental uncertainty, and have considered the
depinning frequency as parameter. The lines of
Figs.~\ref{Tintermed} and \ref{TnearTc} have been obtained by this
procedure.

\begin{figure}
 \includegraphics[width=8cm]{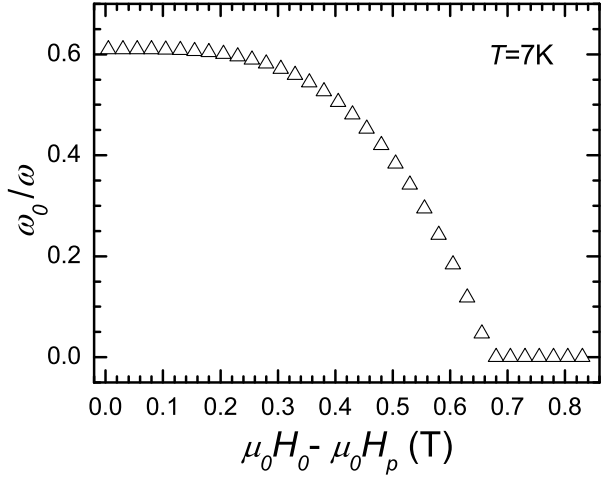}
  \caption{Magnetic field dependence of the depinning frequency, obtained for sample P6 by
  fitting the experimental results reported in Fig.~\ref{Tintermed}b.}
  \label{frequenza-depinning}
\end{figure}

By fitting the results at $T=7$~K, we have obtained the field
dependence of $\omega_0/\omega$ reported in
Fig.~\ref{frequenza-depinning}. The roughly constant value of
$\omega_0/\omega$ we obtained up to $\mu_0 H_0\approx0.3$~T
indicates that in this field range individual vortex pinning
occurs; on further increasing the magnetic field, the interaction
between fluxons becomes important, collective vortex pinning sets
in and, consequently, the depinning frequency decreases. The data
obtained at $\mu_0 H_0 \gtrsim 0.65$~T are well fitted setting
$\omega_0/\omega=0$ in Eq.~(\ref{delta-v(omega)}); this means
that, at $T=7$~K and $\mu_0 H_0 \gtrsim 0.65$~T, the induced mw
current makes fluxons move in the flux-flow regime.

The best-fit curve of Fig.~\ref{TnearTc} has been obtained with
$\omega_0/\omega=0$, as expected. Indeed, at temperature very
close to $T_c$, the pinning effects are weak and the induced mw
current makes fluxons move in the flux-flow regime.

\subsection{Field dependence of $R_s$ in sample P0}
It has been shown by several authors that the properties of the
two-gap \mgb\ superconductor in the mixed state cannot be
accounted for by standard
theories~\cite{sologu,bouquet-H,Magnetiz,Jia,Yang,shibata,dulcic,Sarti}.
It is by now accepted that this is related to the double-gap
nature of \mgb\ that is responsible for an unusual vortex
structure. Indeed, it has been highlighted, both experimentally
and theoretically, that the vortex cores are characterized by two
different spatial and magnetic-field
scales~\cite{koshelev,eskil}. Because of the different
magnetic-field dependence of the two gaps, on varying the field,
the structure of the vortex lattice changes in an unusual way. At
low magnetic fields, quasiparticles by $\pi$ and $\sigma$ bands
are trapped within the vortex core, even if on different spatial
scales because of the different coherence lengths $\xi_{\pi}$ and
$\xi_{\sigma}$; in the field range $0.5 \div 1$~T, though
$\sigma$-band quasiparticles remain localized, the $\pi$-band
quasiparticles spread over the sample~\cite{eskil}; on further
increasing the field, $\Delta_{\pi}$ is strongly reduced, the
$\pi$-quasiparticle contribution remains almost unchanged while
the $\sigma$-quasiparticle contribution continues to increase with
about the same rate up to the macroscopic $H_{c2}$. So, a further
characteristic field is needed for determining the fluxon-lattice
properties of \mgb; the existence of this crossover field, often
indicated as $H_{c2}^{\pi}$, has been highlighted in several
experiments~\cite{sologu,bouquet-H,eskil,cubit,samuely,gonnelli1}.
The field-induced evolution of the vortex lattice is expected to
affect the vortex-vortex and vortex-pinning interactions, making
the standard models most likely inadequate to describe the fluxon
dynamics.
%%%%%%%%%%%%%%

Results on the field-induced variations of $R_s$ in \mgb\ have been reported by some authors~\cite{shibata,dulcic,nova,isteresiMgB2,Sarti,EUCAS2007,zait}; most of them have highlighted several anomalies, which cannot be explained in the framework of standard models for fluxon
dynamics. In particular, it has been highlighted unusually enhanced field-induced mw losses at applied magnetic field much lower than $H_{c2}$. Only Zaitsev et \emph{al.}~\cite{zait} have explained the frequency and field dependence of the mw surface resistance of \mgb\ films in the framework of standard models.
The results we have obtained in sample P0 are similar to those reported by
Shibata et \emph{al.}~\cite{shibata}, who investigated the field
dependence of the surface impedance in \mgb\ single crystal in a wide
range of DC magnetic fields (up to 14~T). At low temperatures, the authors
have observed an initial fast variation of the field-induced mw
dissipation up to fields of the order of 1~T, followed by a slower one at
higher fields. Consistently with the sharp field-induced variation of the
heat capacity and thermal conductivity, the enhanced low-field variation
of the mw losses has been ascribed to the high increase of $\pi$
quasiparticles in the vortex cores. At higher fields, the variation is
slower because of the saturation of the $\pi$-quasiparticle contribution.

According to Shibata et \emph{al.}, the enhanced field-induced
variation we observed in sample P0 can be qualitatively ascribed
to the strong reduction of $\Delta_{\pi}$ in the field range we
have investigated. However, the observed $R_s(H_0)$ curves differ
from the expected ones in both the intensity and the shape. Here,
we discuss the shape of the $R_s(H_0)$ curves.

In a wide range of temperatures below $T_c$, we have observed a
magnetic hysteresis, which should be related to the different
magnetic induction at increasing and decreasing fields, due to the
critical state. As discussed in Sec.~\ref{model}, the
increasing-field branch of the $R_s(H_0)$ curve should exhibit a
change of concavity, from positive to negative, when the external
magnetic field reaches the full penetration field, $H^*$. By
considering the sample width and the value of $J_c$ at $T=5$~K
reported for sample P0~\cite{tarantini1}, the expected value of
$H^*$ is $\approx 2.6$~T. Nevertheless, we observe a negative
concavity of the $R_s(H_0)$ increasing-field branch in the whole
range of fields investigated (see Fig.~\ref{T4K}a), even if the
maximum value of the applied field is well below $H^*$. The
decreasing-field branch should show a monotonic reduction of $R_s$
down to low fields. In contrast, for $H_0<H^{\prime}$, we observe
an initial weak reduction followed by a plateau, from $\mu_0 H_0
\approx 0.2$~T down to zero. The presence of this plateau is
puzzling because it would suggest that the trapped flux does not
change anymore on decreasing the field below $\sim 0.2$~T,
although this value is four times larger that $H_p$.

Another anomalous result concerns the range of magnetic fields in
which we observe the hysteretic behavior. As we have already
mentioned, we have experienced that for samples of millimetric
size the sensitivity of our experimental apparatus allows
detecting hysteresis in $R_s(H_0)$ for $J_c\gtrsim 10^4
\mathrm{A/cm}^2$. From Fig.~9 of Ref.~\cite{tarantini1}, one can deduce that, in
sample P0, such condition occurs at $\mu_0 H_0 \sim 4$~T; so, we
should detect hysteresis in the whole range of fields we have
investigated. On the contrary, we obtained
$H^{\prime}(4.2~\mathrm{K})\sim 0.5$~T, one order of magnitude
lower than the expected value.

We would like to remark that these anomalies have been observed in all the
bulk \mgb\ samples (unirradiated) we have investigated, no matter
the preparation method and the components ($^{11}$B or $^{10}$B)
used in the synthesis process~\cite{isteresiMgB2,EUCAS2007}. The
finding that in sample P6 the experimental results are fully
justified by the used model, strongly suggests that these
anomalies are strictly related to the presence of the two
superconducting gaps.

At temperatures close to $T_c$ and for $H_0 \gtrsim 0.5H_{c2}(T)$,
the experimental results can be accounted for by the model
discussed in Sec.~\ref{model}, provided that the anisotropy of the
upper critical field is taken into due account. Following Ref.~\cite{pallecchi},
to take into account the anisotropy, we have assumed that the
polycrystalline sample is constituted by grains with the $c$-axis
randomly oriented with respect to the DC-magnetic-field direction;
so, the distribution of their orientations follows a
$\sin(\theta)$ law, being $\theta$ the angle between
$\emph{\textbf{H}}_0$ and $\mathcal{\mathbf{\hat{c}}}$.
Furthermore, we have used for the angular dependence of the upper
critical field the anisotropic Ginzburg-Landau relation
\begin{equation*}
    H_{c2}(\theta) = \frac{H_{c2}^{\perp
    c}}{\sqrt{\gamma^2 \cos^2(\theta) + \sin^2(\theta)}} \,,
\end{equation*}
where $\gamma = H_{c2}^{\perp c}/H_{c2}^{\parallel c}$ is the
anisotropy factor.

The field-induced variations of $R_s$ observed at $T=38$~K
(reported in Fig.~\ref{TnearTc}a) do not exhibit hysteresis; so,
in this case, $B$ can be considered uniform. Furthermore, at
temperatures near $T_c$, one can reasonably suppose fluxons move
in the flux-flow regime. In this condition, the expected $R_s(H_0,
H_{c2}(\theta))$ curve depends on $\lambda_0/\delta_0$ (obtained
by fitting the $R_s(T)$ curve at $H_0=0$), $H_{c2}^{\perp c}$ and
$\gamma$. On the other hand, the $H_{c2}$ values deduced from the
isothermal $R_s(H_0)$ curves (see Fig.~\ref{campicaratteristici}b)
coincide with the magnetic field at which the whole sample goes to
the normal state, i.e. $H_{c2}^{\perp c}$. In order to fit the
results at $T=38$~K, we have averaged the expected curve
[calculated by Eqs.~(\ref{Rs}--\ref{delta0})] over a
$\sin(\theta)$ distribution, have used for $H_{c2}^{\perp c}$ the
value of the magnetic field at which $R_s/R_n=1$, letting it vary
within the experimental uncertainty, have taken $\gamma$ as free
parameter. At this value of temperature, the experimental results
can be accounted for using $\gamma=3.3 \pm 0.5$. In particular,
the best-fit curve reported in Fig.~\ref{TnearTc}a has been
obtained with $\gamma=3.3$ and $\mu_0H_{c2}^{\perp c}=145$~mT. As
one can see, the field-induced variation of $R_s$ is well
described by the model of Sec.~\ref{model}. We think that this
occurs because at this temperature the superfluid fraction of the
$\pi$ band is strongly suppressed at low magnetic fields, the flux
line gets a conventional structure and the fluxon dynamics can be
described by standard models.
\begin{figure}[ht]
 \includegraphics[width=8cm]{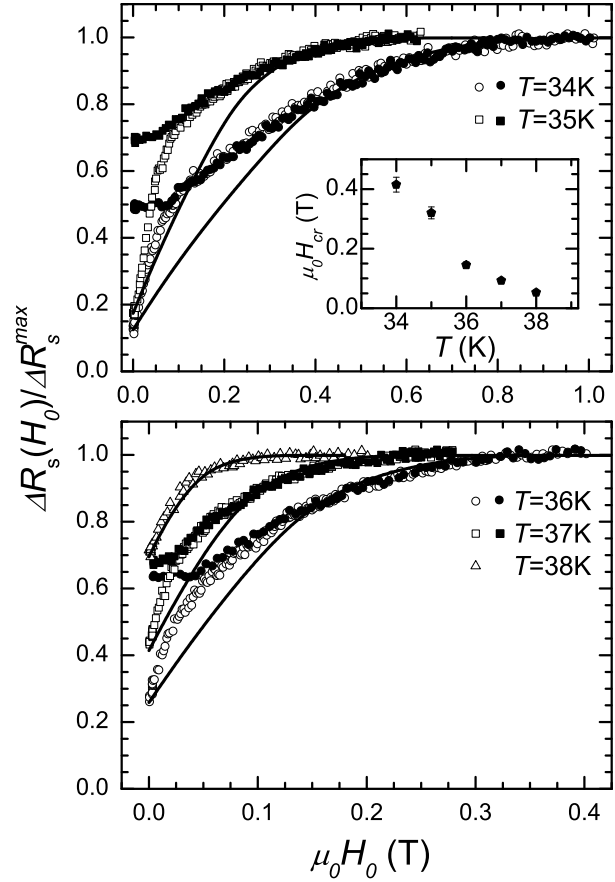}
  \caption{Normalized field-induced variations of $R_s$ for sample P0, at
  different temperatures near $T_c$. Open symbols are the
results obtained at increasing $H_0$, full symbols those at
decreasing $H_0$. The $R_s(H_0)$ curve at $T=38$~K is reversible.
The lines are the expected curves, obtained with $\gamma=2.6$ as
described in the text. The inset shows the temperature dependence
of the magnetic field $H_{cr}$, above which the experimental data
can be justified in the framework of the model of
Sec.~\ref{model}.} \label{Fit-near-Tc}
\end{figure}

Prompted by the results obtained at $T=38$~K, we have tried to fit
the experimental data obtained in the temperature range
$34\div37$~K by the same method (for these temperatures, the upper
critical field has been directly deduced from the $R_s(H_0)$
curves). Since in this temperature range we have detected magnetic
hysteresis, we have considered only the reversible part of the
$R_s(H_0)$ curve. In order to fit the data, we have hypothesized
fluxons move in the flux-flow regime, have considered for
$H_{c2}^{\perp c}(T)$ the values reported in the inset of
Fig.~\ref{campicaratteristici}b, letting them vary within the
experimental uncertainty, and have taken $\gamma$ as fitting
parameter. We have found that at high fields the experimental
results can be fitted using $\gamma=2.6\pm0.2$.
Fig.~\ref{Fit-near-Tc} shows a comparison between the expected
curves, obtained with $\gamma=2.6$, and the experimental data for
$T=34\div38$~K; open symbols are the results obtained at
increasing $H_0$, full symbols those at decreasing $H_0$. As one
can see, the expected $R_s(H_0)$ curve at $T=38$~K so obtained
poorly agrees with the experimental data at low fields; on the
contrary, the line of Fig.~\ref{TnearTc}a, which has been obtained
with $\gamma=3.3$, fits the data in the whole range of magnetic
fields. However, the value $\gamma=2.6$ is closer to the upper-critical-field anisotropy reported in the
literature for \mgb\ at temperatures near $T_c$~\cite{golubovHc2,cubit,caplin,Lyard}.

The results of Fig.~\ref{Fit-near-Tc} show that for $H_0$ greater
than a certain threshold value, depending on $T$, the data can be
justified in the framework of the model describing the fluxon
dynamics of conventional vortex lattice. The temperature
dependence of the threshold field, $H_{cr}$, is reported in the
inset. For $H_0 < H_{cr}$ the field-induced mw losses are larger
than those expected for single-gap superconductors in the mixed
state. We suggest that this surplus of mw losses is due to the
additional contribution of $\pi$-band quasiparticles within the
vortex cores with respect to that of one-gap superconductors; the
finding that $H_{cr}$ decreases on increasing $T$ seems to support
this hypothesis. It is easy to see that the $H_{cr}(T)$ values
coincide, within the experimental uncertainty, with $0.5H_{c2}(T)$
(see the inset of Fig.~\ref{campicaratteristici}). Presently, it
is not clear why just above $0.5H_{c2}$ the results can be
justified by a standard model, which does not consider the two-gap
nature of \mgb. Furthermore, we remark that $H_{cr}$ cannot be
identified with the magnetic field at which the $\pi$-band
superfluid is suppressed; indeed several authors have reported
$H_{c2}^{\pi}\sim 0.1H_{c2}^{\perp
c}$~\cite{sologu,bouquet-H,eskil,cubit,samuely,gonnelli1}.

\section{Conclusions}
We have investigated the microwave surface resistance at 9.6~GHz of two
polycrystalline $\mathrm{Mg}^{11} \mathrm{B}_2$ samples prepared by direct
synthesis from Mg (99.999\% purity) and crystalline isotopically enriched
$^{11}$B (99.95\% purity). That labelled as P0 consists of pristine
material; the other, labelled as P6, has been exposed to neutron
irradiation at very high fluence. Several superconducting properties of
these samples have been reported in Refs.~\cite{tarantini1,pallecchi,gonnelli2,putti2,puttiSUST2008}. Point-contact
spectroscopy and specific-heat measurements, have shown that sample P0
exhibits a clear two-gap-supercon-ductivity behavior; in sample P6 the
irradiation process determined a merging of the two gaps into a single
value. To our knowledge, the mw response of
neutron irradiated \mgb\ samples has not yet been investigated.

The mw surface resistance has been measured as a function of the
temperature and the DC magnetic field. By measuring the
field-induced variations of $R_s$ at increasing and decreasing
fields we have detected a magnetic hysteresis ascribable to the critical state of the fluxons lattice. The range of temperatures in which the hysteretic behavior has been observed is
different for the two samples; in the irradiated sample the
hysteresis is undetectable at $T/T_c \gtrsim 0.55$ while in the
unirradiated sample it is detectable up to $T/T_c \approx 0.95$.

The results obtained in the irradiated sample have been quite well justified
in the framework of the Coffey and Clem model with the normal
fluid density following the Gorter and Casimir two-fluid model. In
order to account for the hysteretic behavior, we have used a
generalized Coffey and Clem model in which we take into account
the non-uniform fluxon distribution due to the critical state.

The peculiarities of the mw surface resistance of sample P0 differ
from those observed in sample P6, in both the temperature and the
field dependencies. The $R_s(T)$ curve obtained at zero field
shows a wide tail, from $T/T_c\approx 0.9$ down to $T/T_c \approx
0.7$, which cannot be justified in the framework of the Gorter and
Casimir two-fluid model. We have shown that, in order to account
for this behavior, it is essential to hypothesize a linear
temperature dependence of the normal and condensed fluid
densities. Such finding agrees with the experimental temperature
dependence of the penetration depth reported in the literature,
which have been justified in the framework of two-gap models for
the MgB$_2$ superconductor.

The $R_s(H_0)$ curves in sample P0 have shown several anomalies,
especially at low temperatures, among which an enhanced
field-induced variation and a magnetic hysteresis of
unconventional shape. At low temperatures, a magnetic field
$H_0\approx H_{c2}/15$ causes a $R_s$ variation of $\approx 35\%$
of the normal-state value. We remark that in sample P6 a variation
of the same order of magnitude is obtained for $H_0 \approx
H_{c2}/2$. The shape of the magnetic hysteresis, which has been
observed in a wide range of temperatures below $T_c$, cannot be
justified in the framework of the critical-state models; the most
unexpected behavior concerns the decreasing-field branch, in which
we observed a plateau extending from $\mu_0 H_0 \sim 0.2$~T down
to zero. The presence of this plateau is puzzling because it would
suggest that the trapped flux does not change anymore on
decreasing the field below 0.2~T, although this value is four
times larger than the first penetration field.

The investigation at temperatures near $T_c$ has highlighted that,
in the range $T=34\div 38$~K, the results obtained in sample P0
for $H_0\gtrsim 0.5H_{c2}$ can be justified in the framework of
the Coffey and Clem model taking into account the anisotropy of
the upper critical field. We suggest that this occurs because at
these field values the superfluid fraction of the $\pi$ band is
strongly suppressed, the flux line gets a conventional structure
and the fluxon dynamics can be described by standard models.

The enhanced field-induced variation of $R_s$, observed at low $T$
in the whole range of fields investigated as well as at $T \sim
T_c$ for $H_0 \lesssim 0.5H_{c2}$, may be qualitatively ascribed
to the presence and motion of the giant cores due to the
$\pi$-band quasiparticles. On the contrary, the origin of the
anomalous shape of the $R_s(H_0)$ curve is so far not understood.
We would like to remark that the results we obtained in sample P0 are very
similar to those, not reported here, we have obtained in several
\mgb\ samples (unirradiated), no matter the preparation method and
the components ($^{11}B$ or $^{10}B$) used in the synthesis
process. The comparison between the results obtained in the two
samples here investigated strongly suggest that the anomalies in
the $R_s(H_0)$ curves are related to the unusual structure of
fluxons due to the two superconducting gaps. According to what
suggested by different authors, our results confirm that the
standard models are inadequate to describe the fluxon dynamics in
two-gap \mgb. Further investigation is necessary for understanding
how to take into account the complex vortex structure in
describing the fluxon dynamics in \mgb.

\section*{Acknowledgements}
The authors are very glad to thank D. Daghero, G. Ghigo, R. S. Gonnelli and
M. Putti for their interest to this work and helpful suggestions;
G. Lapis and G. Napoli for technical assistance.

%

%\begin{flushright}\today \end{flushright}

\end{document}